\documentclass[conference,letter]{IEEEtran}

\usepackage{amsmath,amsthm,amssymb,amsfonts,xpatch}
\usepackage{array}
\usepackage[acronym,toc]{glossaries}
\usepackage[table,xcdraw]{xcolor}
\usepackage{subcaption}
\usepackage{mathtools}
\usepackage{todonotes}
\usepackage{dirtytalk}
\usepackage{nicefrac}
\usepackage{multirow}
\usepackage{textcomp}
\usepackage{booktabs}
\usepackage{textcomp}
\usepackage{graphicx}
\usepackage{latexsym}
\usepackage{paralist}
\usepackage{relsize}
\usepackage{balance}
\usepackage{comment}
\usepackage{pifont}
\usepackage{xfrac}
\usepackage{tasks}
\usepackage{cite}
\usepackage{enumerate}
\usepackage{xspace}
\usepackage{stfloats} 
\usepackage{fontawesome5}
\usepackage{hyphenat}
\usepackage{booktabs}
\usepackage{multirow}
\usepackage{listings}
\usepackage{url}
\lstdefinelanguage{yaml}{
  keywords={attack_params, attack_type, args, query_budget, benign_query_percentage, epsilon_mifgsm, distributed_attack_manager, num_clients, coordination_strategy, phases, initial_benign_queries, type, num_fixed_queries, attack, post_attack_benign_queries},
  keywordstyle=\color{blue!40!black},
  morestring=[b]",
  stringstyle=\color{green!50!black},
  morecomment=[l]{\#},
  commentstyle=\color{gray}\itshape,
  alsoletter={_}, % Wichtig für Keywords mit Unterstrich
  sensitive=true,
  basicstyle=\ttfamily\scriptsize, % Kleiner, um Platz zu sparen
  numbers=left,
  numberstyle=\tiny\color{gray},
  stepnumber=1,
  breaklines=true,
  frame=single,
  tabsize=2,
  xleftmargin=2em,
  framexleftmargin=1.5em
}
\usepackage{xcolor}
\usepackage{algorithm}
\usepackage{algpseudocode}
\makeatletter
\renewcommand{\theALG@line}{0}
\makeatother

\usepackage[letterpaper, left=0.625in, right=0.625in, bottom=1.1in, top=0.75in]{geometry}

\colorlet{punct}{red!60!black}
\definecolor{background}{HTML}{EEEEEE}
\definecolor{delim}{RGB}{20,105,176}
\colorlet{numb}{magenta!60!black}

\lstdefinelanguage{json}{
    basicstyle=\ttfamily\scriptsize,
    numbers=left,
    numberstyle=\scriptsize,
    stepnumber=1,
    numbersep=2pt,
    showstringspaces=false,
    breaklines=true,
    frame=lines,
    backgroundcolor=\color{background},
    literate=
     *{0}{{{\color{numb}0}}}{1}
      {1}{{{\color{numb}1}}}{1}
      {2}{{{\color{numb}2}}}{1}
      {3}{{{\color{numb}3}}}{1}
      {4}{{{\color{numb}4}}}{1}
      {5}{{{\color{numb}5}}}{1}
      {6}{{{\color{numb}6}}}{1}
      {7}{{{\color{numb}7}}}{1}
      {8}{{{\color{numb}8}}}{1}
      {9}{{{\color{numb}9}}}{1}
      {:}{{{\color{punct}{:}}}}{1}
      {,}{{{\color{punct}{,}}}}{1}
      {\{}{{{\color{delim}{\{}}}}{1}
      {\}}{{{\color{delim}{\}}}}}{1}
      {[}{{{\color{delim}{[}}}}{1}
      {]}{{{\color{delim}{]}}}}{1},
}
\definecolor{Gray}{gray}{0.95}

\makeatletter
   \xpatchcmd{\@thm}{\fontseries\mddefault\upshape}{}{}{} % same font as thm-header
\makeatother

\def\BibTeX{{\rm B\kern-.05em{\sc i\kern-.025em b}\kern-.08em T\kern-.1667em\lower.7ex\hbox{E}\kern-.125emX}}
\theoremstyle{definition}

\definecolor{promptSystem}{RGB}{102,102,255}    % Blue for system prompts
\definecolor{promptUser}{RGB}{51,153,102}      % Green for user prompts
\definecolor{promptAssistant}{RGB}{255,102,102}% Red for assistant prompts
\definecolor{promptVariable}{RGB}{255,153,51}  % Orange for variables
\definecolor{promptName}{RGB}{153,51,255}      % Purple for prompt names
\definecolor{grayDark}{RGB}{50,50,50}          % Dark gray for text
\definecolor{grayMedium}{RGB}{70,70,70}        % Medium gray
\definecolor{grayLight}{RGB}{90,90,90}         % Light gray

\lstdefinelanguage{LLMPrompt}{
  morekeywords={Prompt,Name}, % Keywords for prompt name
  morekeywords=[2]{System}, % Keywords for system role
  morekeywords=[3]{User}, % Keywords for user role
  morekeywords=[4]{Assistant}, % Keywords for assistant role
  sensitive=true, % Case-sensitive
  morecomment=[l]{@/**}, % Line comment starting with @/**
  morecomment=[s]{@/**}{**/@}, % Start and end comment delimiters
  morestring=[b]", % Strings enclosed in double quotes
  alsoletter={\{}{\}}, % Treat curly braces as special characters
  moredelim=[is][\color{promptVariable}]{\{}{\}}, % Highlight variables in {}
}

\begin{document}
        % Header
        \title{AI Model Extraction Attacks: Bypassing Single-Client Assumptions in Defenses}

%\title{AI Model Extraction Attacks: Bypassing Single-Client Assumptions in Defences}

%\title{Bypassing Single-Client Assumptions in Defences for AI Model Extraction Attacks}

%\title{The Myth of a Single Attacker: Bypassing Single-Client Assumptions in MEA Defenses}

        \author{
  \IEEEauthorblockN{
        Maxime Schwarzer $^{1,2}$, Johannes F. Loevenich$^{1,3}$, Gustavo Sánchez$^{2}$, Laurin Holz$^{1,4}$,\\ Thies Möhlenhof$^{1,5}$, Tobias Hürten$^{1}$, Roberto Rigolin F. Lopes$^{1}$, and Veit Hagenmeyer$^{2}$}  
  \IEEEauthorblockA{
        $^{1}$CortAIx Labs, Thales Deutschland, Ditzingen, Germany \\        
        $^{2}$Institute for Automation and Applied Informatics (IAI), Karlsruhe Institute of Technology (KIT), Karlsruhe, Germany \\
        $^{3}$Department of Mathematics/Computer Science, University of Osnabrück, Osnabrück, Germany \\        
        $^{4}$Department of Computer Science, University of Ulm, Ulm, Germany \\
        $^{5}$Department of Computer Science, University Koblenz-Landau, Koblenz, Germany \\
        Email: \{maxime.schwarzer, johannes.loevenich, roberto.rigolin\}@thalesgroup.com, \{sanchez, veit.hagenmeyer\}@kit.edu        
  }
}

        \maketitle
        \IEEEpubidadjcol
        % Double blind review
        %\input{header/03_Author}        

        % Acronyms
        %% LaTeX2e class for student theses
%% 004acronyms.tex
%%!TEX root = ./thesis.tex
%%
%% Karlsruhe Institute of Technology
%% Institute for Automation and Applied Informatics (IAI)
%% IT Methods and Components for Energy Systems (IT4ES) Research Group
%%
%% Gökhan Demirel
%% goekhan.demirel@kit.edu
%%
%% Adaptation Version 1.1, 19.02.2024

\newacronym{ACD}{ACD}{Autonomous Cyber Defence}
\newacronym{IAI}{IAI}{Institute for Automation and Applied Informatics} % \gls{IAI}
\newacronym{BES}{BES}{battery energy storage} % \gls{BES}
\newacronym{DER}{DER}{distributed energy resources} % \gls{DER}
\newacronym{p.u}{$p.u$}{per unit} % \gls{p.u}
\newacronym{VSCs}{VSCs}{voltage source converters} % \gls{VSCs}
\newacronym{DSO}{DSO}{Distribution System Operator} % \gls{DSO}
\newacronym{RMS}{RMS}{root mean square} % \gls{RMS}
\newacronym{AV}{AV}{Actual Value} % \gls{AV}
\newacronym{TV}{TV}{Target Value} % \gls{TV}
\newacronym{C4ISR}{C4ISR}{Command, Control, Communications, Computers, Intelligence, Surveillance, and Reconnaissance}
\newacronym{SGs}{SGs}{Smart Grids}
\newacronym{ML-based IDSs}{ML-based IDSs}{ Machine Learning-based Intrusion Detection Systems}
\newacronym{DNN}{DNN}{Deep Neural Networks}
\newacronym{FDI}{FDI}{Feature Distortion Index}
\newacronym{ML}{ML}{Machine Learning}
\newacronym[plural=MEAs]{MEA}{MEA}{Model Extraction Attack}
\newacronym{MLSys}{ML-based systems}{machine learning-based systems}
\newacronym{MLaaS}{MLaaS}{ML-as-a-Service}
\newacronym{FGSM}{FGSM}{Fast Gradient Sign Method}
\newacronym{CWA}{C\&W attack}{Carlini \& Wagner attack}
\newacronym{PGD}{PGD}{Projected Gradient Descent}
\newacronym{ZOO}{ZOO}{Zeroth-Order Optimization}
\newacronym{SOTA}{SOTA}{state-of-the-art}
\newacronym{SDN}{SDN}{Software-Defined Networking}
\newacronym{DRL}{Deep RL}{Deep Reinforcement Learning}
\newacronym{GQM}{GQM}{Goal-Question-Metric}
\newacronym{AI}{AI}{Artificial intelligence}
\newacronym{MTS}{MTS}{Multivariate time series}
\newacronym{NLP}{NPL}{Natural Language Processing}
\newacronym{AI}{AI}{Artificial Intelligence}
\newacronym[plural=AICAs]{AICA}{AICA}{Autonomous Intelligent Cyber-defence Agent}
\newacronym{AGI}{AGI}{Artificial General Intelligence}
\newacronym{API}{API}{Application Programming Interface}
\newacronym{APIs}{APIs}{Application Programming Interfaces}
\newacronym[plural=APTs]{APT}{APT}{Advanced Persistent Threat}
\newacronym{AWS}{AWS}{Amazon Web Services}
\newacronym{AUC}{AUC}{Area Under the Curve}
\newacronym{DL}{DL}{Deep Learning}
\newacronym{YAML}{YAML}{YAML Ain’t Markup Language}
\newacronym[plural=EDFs]{EDF}{EDF}{Empirical Distribution Function}

\newacronym{SGD}{SGD}{Stochastic Gradient Descent}
\newacronym{GAN}{GAN}{Generative Adversarial Network}

\newacronym{IaC}{IaC}{Infrastructure as Code}

\newacronym[plural=IDSs]{IDS}{IDS}{Intrusion Detection Systems}
\newacronym{ID}{ID}{Intrusion Detection}
\newacronym{IPS}{IPS}{Intrusion Prevention System}

\newacronym{IP}{IP}{Intellectual Property}

\newacronym[plural=LLMs]{LLM}{LLM}{Large Language Model}
\newacronym{LSTM}{LSTM}{Long Short-Term Memory}
\newacronym[plural=SVMs]{SVM}{SVM}{Support Vector Machine}
\newacronym{MLP}{MLP}{Multilayer Perceptron}
\newacronym{ML}{ML}{Machine Learning}
\newacronym{SML}{SML}{Shallow Machine Learning}
\newacronym{NN}{NN}{Neural Network}

\newacronym{kNN}{kNN}{k-Nearest Neighbors}
\newacronym{RL}{RL}{Reinforcement Learning}
\newacronym{APT}{APT}{Advanced Persistent Threat}
\newacronym{APTs}{APTs}{advanced persistent threats}
\newacronym{SDD}{SDD}{Software Defined Defence}
\newacronym{SDN}{SDN}{Software Defined Networking}
\newacronym{APMSA}{APMSA}{Adversarial Perturbation Against Model Stealing Attacks}
\newacronym{MLIDS}{MLIDS}{Machine learning-based intrusion detection systems}
\newacronym{I-FGSM}{I-FGSM}{Iterative Fast Gradient Sign Method}
\newacronym{MI-FGSM}{MI-FGSM}{Momentum Iterative Fast Gradient Sign Method}
\newacronym{PRADA}{PRADA}{Protecting Against DNN Model Stealing Attacks}
\newacronym{ADAM}{ADAM}{Adaptive Moment Estimation}
\newacronym{RQs}{RQs}{Research Questions}
\newacronym{NATO}{NATO}{North Atlantic Treaty Organization}

\newacronym{CNN}{CNN}{Convolutional Neural Network}
\newacronym{FCN}{FCN}{Fully-Connected Feed Forward Neural Network}
\newacronym{LLMs}{LLMs}{large language models}
\newacronym{CV}{CV}{Computer Vision}
\newacronym{RAC}{RAC}{Relational Attention Corrector}
\newacronym{MMD}{MMD}{Maximum Mean Discrepancy}
\newacronym{KL divergence}{KL divergence}{Kullback-Leibler divergence}
\newacronym{PCA}{PCA}{Principal Component Analysis}

\newacronym{CTGAN}{CTGAN}{Conditional Generative Adversarial Network}
\newacronym{ERENO}{ERENO}{Efficacious Reproducer Engine for Network Operations}
\newacronym{RKHS}{RKHS}{Reproducing Kernel Hilbert Space}
\newacronym{SCA}{SCA}{Single Client Assumption}
\newacronym{TP}{TP}{True Positives}
\newacronym{FP}{FP}{False Positives}
\newacronym{TN}{TN}{True Negatives}
\newacronym{FN}{FN}{False Negatives}
\newacronym{ROC}{ROC}{Receiver Operating Characteristic}
\newacronym{TPR}{TPR}{True Positive Rate}
\newacronym{FPR}{FPR}{False Positive Rate}
\newacronym{AUC}{AUC}{Area Under the Curve}
% Define glossary entries
\newglossaryentry{backpropagation}{name=Backpropagation, text=backpropagation, description={Propagating gradients obtained from an error function in backward direction through a network}}
\newglossaryentry{latex}{
    name=\LaTeX,
    description={A document preparation system}
}

\newglossaryentry{example}{
    name=Example,
    description={An example entry}
}

\newacronym{QUEEN}{QUEEN}{Query Unlearning}
    
        % Sections
        % The abstract should not exceed 1000 characters
\begin{abstract}
Ensuring the protection of Artificial Intelligence (AI) models deployed in military Command and Control (C2) systems and critical infrastructure is essential for maintaining information superiority. Model Extraction Attacks (MEAs) pose a significant threat, as they enable adversaries to replicate proprietary models, compromise protected information, and prepare offline adversarial attacks. However, current defense strategies predominantly rely on the Single Client Assumption (SCA), which is the implicit assumption that attacks originate from isolated identities. This work systematically demonstrates that the SCA is fundamentally invalid in the presence of coordinated threat actors, such as Advanced Persistent Threats (APTs). We introduce a modular, open-source framework called CerberusAI for reproducible model-stealing research, and use it to simulate distributed attack scenarios. Our empirical evaluation shows that well-established defense mechanisms, such as Protecting Against Deep Neural Network Model Stealing Attacks (PRADA), can be bypassed by basic round-robin query distribution strategies, resulting in a significant reduction in detection performance. Furthermore, we demonstrate that even global aggregation approaches can be rendered operationally useless through adaptive \emph{traffic mixing}. These results highlight the need for a paradigm shift towards stateful, identity-independent defense architectures in the field of model extraction attacks. This paper was originally presented at the International Conference on Military Communication and Information Systems (ICMCIS), organized by the Information Systems Technology (IST) Scientific and Technical Committee, IST-224-RSY – the ICMCIS, held in Bath, United Kingdom, 12-13 May 2026.

% Needed for submission see (https://icmcis.eu/wp-content/uploads/2025/08/20250807_UC_ICMCIS2026_CfP.pdf): ICMCIS is open for scientific papers at PUBLIC RELEASE level, to be submitted in the IEEE -Manuscript Templates for Conference Proceedings, with the addition of the following in the abstract:
% This paper was originally presented at the International Conference on Military Communication and Information Systems (ICMCIS), organized by the Information Systems Technology (IST) Scientificand Technical Committee, IST-224-RSY – the ICMCIS, held in Bath, United Kingdom, 12-13 May2026
\end{abstract}

        \section{Introduction} 
\label{sec:introduction}

Modern military units and operators of critical infrastructure increasingly rely on \gls{AI} models as integral components of mission-critical systems, ranging from cyber defense platforms and energy grid management to tactical \gls{C4ISR} services and autonomous sensing architectures~\cite{Loevenich:2025:Journal, Loevenich:ICMCIS2025}. Recent investigations on \gls{SDD}~\cite{soare2023software} identify software as a primary capability driver for future multi-domain operations. As a consequence, military systems must be rapidly adaptable, resilient, and securely updatable across their entire lifecycle. In such highly connected, data-driven operational environments, \gls{AI} models become both strategic enablers and attractive targets for adversaries.

A critical yet often underestimated threat to these models is the \gls{MEA}, in which malicious network nodes attempt to replicate the target \gls{AI} model. Through systematic querying, an adversary can reconstruct a proprietary model's decision boundary~\cite{orekondy2018knockoff}, enabling the theft of sensitive confidential information and facilitating downstream attacks such as adversarial evasion of deployed \gls{IDS}~\cite{papernotAdvExampleFooling}. Within military contexts, such model extractions directly undermine the \gls{SDD} goals of rapid capability evolution and secure deployment pipelines by exposing the underlying \gls{AI} components to manipulation or replication.

Current \gls{MEA} defense mechanisms, however, rely heavily on the \gls{SCA}, i.e., the hypothesis that malicious activity originates from a single, isolated client. Defenses such as \gls{PRADA}~\cite{prada}, \gls{QUEEN}~\cite{queen} and Model-Guardian~\cite{ModelGuardian} operate by identifying anomalous request patterns on a per-client basis. Yet the \gls{SDD} paradigm explicitly anticipates highly networked, federated environments with diverse clients, shared middleware layers, and modular \gls{AI} services distributed across platforms. In such settings, whether in a tactical edge network or a distributed smart city architecture, adversaries with sufficient resources can trivially distribute their probing across many coordinated clients, rendering \gls{SCA}-based defenses ineffective.

This discrepancy between theoretical assumptions and operational threat realities highlights a structural security gap: current defenses do not align with the security requirements of detecting coordinated, adaptive attack patterns in globally aggregated data streams. Addressing this gap is essential for enabling trustworthy \gls{AI} components in \gls{SDD} architectures, particularly where \gls{AI} agents support time-critical decisions, \gls{ACD}, or cross-platform analytics in federated defense and infrastructure networks. Therefore, this paper discusses the fundamental flaw in the underlying assumption itself and introduce an open-source framework for evaluating realistic, distributed, and adaptive \gls{MEA} threats.

Our main contributions are:

\begin{enumerate}
    \item \emph{Systematic demonstration of \gls{SCA} vulnerability}: Using \gls{PRADA} as a representative mechanism, we show how coordinated, distributed query strategies reduce detection performance to zero.
    \item \emph{Analysis of the limits of global defenses}: We demonstrate that even naive global request aggregation fails once attackers embed malicious probes into legitimate background traffic, mirroring mixed operational traffic in \gls{SDD}-enabled mission networks.
    \item \emph{Introduction of the open-source framework}: The modular, open-source framework CerberusAI for simulating distributed, adaptive \gls{MEA} scenarios aligned with \gls{SDD}-relevant deployment architectures.
\end{enumerate}

We argue that, by uncovering this structural weakness, future research must move beyond isolated-client analyses and towards robust, learning-enabled security architectures that can recognize advanced forms of coordinated adversarial behavior.

The remainder of this paper is structured as follows: Section~\ref{sec:rationale} highlights the relevance of the topic for military applications and cyber operations. Section~\ref{sec:related_work} provides an overview of related work and the current state of the art. In Section~\ref{sec:sca_fallacy}, we derive the theoretical vulnerability of the single client assumption, after which Section~\ref{sec:framework} introduces the developed CerberusAI framework. Section~\ref{sec:experiments} presents the empirical evaluation of the attack scenarios. The paper concludes with a summary and an outlook on future research areas in Section~\ref{sec:conclusion}.
        \section{Relevance for NATO}
\label{sec:rationale}

The integration of \gls{AI} into \gls{NATO} military capabilities, ranging from autonomous and unmanned systems to decision-support and \gls{IDS} in federated tactical networks, fundamentally reshapes the Alliance’s operational and security landscape. As emphasized in NATO’s Revised \gls{AI} Strategy, \gls{AI} has become a general-purpose and dual-use technology whose military adoption must be accompanied by robust safeguards to ensure reliability, governability, and resilience against adversarial interference. Protecting deployed \gls{AI} models is therefore not merely a technical concern, but a prerequisite for sustaining information superiority, operational trust, and interoperability across Allied forces.

Within this strategic context, \glspl{MEA} represent a critical and under-addressed threat to \gls{NATO}’s \gls{AI} readiness. Beyond the loss of sensitive \gls{IP} and the associated risk of privacy leakage from training data, a successfully extracted model provides adversaries with an operational testbed for systematic exploitation. Possession of a functional replica of a military \gls{AI} component enables the offline development and validation of tailored evasion strategies, allowing \gls{AI}-enabled detection and decision systems to be bypassed during operations without observable interaction with the protected system. Such adversarial use of \gls{AI} directly contradicts NATO’s objectives to protect innovation, manage \gls{AI}-related risks, and ensure the responsible and secure deployment of \gls{AI}-enabled capabilities.

Existing defense approaches for protecting \gls{AI} models largely originate from civilian contexts and implicitly rely on assumptions that do not hold in \gls{NATO}-relevant threat environments. In particular, client-centric anomaly detection mechanisms presuppose isolated attackers and benign operational conditions, whereas NATO must assume coordinated, well-resourced, and adaptive adversaries operating at scale. Advanced Persistent Threats can distribute probing activities across large infrastructures and blend malicious behavior into legitimate traffic, rendering identity-based and purely statistical defenses ineffective. This mismatch between prevailing defense assumptions and realistic adversarial capabilities undermines \gls{NATO}’s ambition to establish a credible Alliance-wide \gls{AI} Testing, Evaluation, Verification, and Validation landscape capable of addressing emerging threats.

By systematically demonstrating the failure of single-client defenses under coordinated attack conditions and by providing a modular framework for realistic red teaming of \gls{AI} systems, this work directly supports \gls{NATO}’s strategic goals of protecting \gls{AI} technologies, strengthening cyber resilience, and improving the Alliance’s understanding of adversarial \gls{AI} use. The presented approach contributes to the development of reproducible testing methodologies aligned with \gls{NATO}’s emphasis on responsible \gls{AI} adoption, threat-informed evaluation, and the protection of \gls{AI}-enabled capabilities throughout their operational lifecycle.
        \section{Related Work} 
\label{sec:related_work}
%This section highlights the current state of the art and identifies the critical gaps addressed by this work.

\subsection{Model Extraction Attacks (MEA)}

Research on \gls{MEA} has developed rapidly in recent years, with a constant arms race between new attack strategies and corresponding defense mechanisms. The goal of \gls{MEA} is to create a functionally equivalent substitute for a proprietary target model by systematically querying its interface. For example, the pioneering work showed early on that even simple query strategies can replicate target models with high precision~\cite{tramèr2016stealingmachinelearningmodels}. To increase efficiency and reduce costs, subsequent work developed active learning strategies that selectively query data points expected to maximize information gain~\cite{ActiveThief, orekondy2018knockoff}.

While early attacks often relied on the availability of similar data sets~\cite{prada}, more recent work on data-free attacks~\cite{maze, tempest} has demonstrated that attackers can be successful even without domain knowledge by using synthetic data generation. This development suggests that the barriers for attackers are continuously decreasing while the complexity of attacks is increasing.

\subsection{Defense Strategies and the Single Client Assumption}

Countermeasures can generally be classified as either passive (e.g., watermarking~\cite{inversionGuidededDefense}) or active. Active methods, which aim to detect attacks at runtime, are particularly critical for safeguarding \gls{AI} models deployed within military communication infrastructures.
\glsreset{PRADA}
The most well-known approach for stateful analysis is \gls{PRADA}~\cite{prada}. This approach analyzes the distribution of the intervals between successive requests from a client and alerts the user in case of deviations from a characteristic normal distribution. Similar approaches such as \gls{QUEEN}~\cite{queen} and Model-Guardian~\cite{ModelGuardian} attempt to disrupt the attacker's gradient estimation by introducing targeted noise into the confidence values. 

A fundamental shortcoming of these approaches is the implicit \gls{SCA}. The detection logic usually operates in isolation for each user ID (API key or IP address). However, as indicated by real incidents and the increasing availability of botnets, coordination across distributed identities is a common practice among resource-rich attackers~\cite{Bederna2020}. Initial approaches such as FDINet~\cite{fdinet} attempt to detect distributed patterns by analyzing internal model activations. However, these methods are often model-specific, not publicly available, and difficult to transfer to black-box scenarios.

\subsection{Lack of Standardized Research Frameworks}

A major bottleneck in the advancement of \gls{MEA} research is the lack of standardized evaluation frameworks that enable meaningful comparison across studies. Experiments are frequently conducted under differing assumptions (threat models, datasets, metrics) and are often implemented using ad hoc scripts.

As S{\'a}nchez et al.~\cite{sanchez2024attacking} point out, there is a lack of standardized open-source tools in the field of learning-based system security, particularly in the context of critical infrastructures such as smart grids that enable systematic evaluation. Existing libraries often focus mainly on adversarial examples (e.g., Adversarial Robustness Toolbox) and do not provide integrated environments for simulating complex, distributed \gls{MEA} scenarios.
           
        \section{The Single Client Assumption}
\label{sec:sca_fallacy}

The prevailing strategy for detecting \gls{MEA} is based on the analysis of query patterns. Methods such as \gls{PRADA} analyze the statistical distribution of queries (e.g., the intervals between consecutive queries) to identify deviations from benign behavior.

These approaches rely on an implicit but critical assumption: the \acrfull{SCA}. In this section, we formalize why this assumption is likely to fail in realistic threat scenarios, particularly in the context of military cyber defense against \glspl{APT}.

\begin{figure}[]
    \centering
    \includegraphics[width=\linewidth]{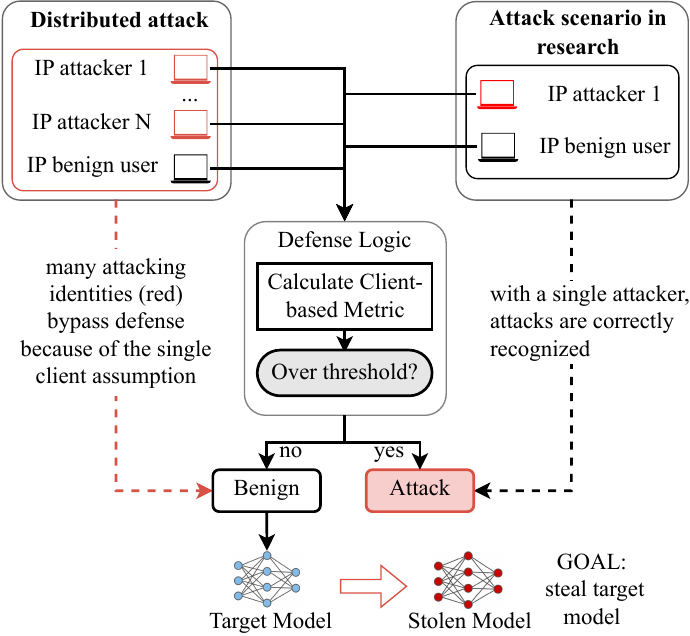} 
    \caption{Visualization of the \gls{SCA} fallacy. Right: Defense recognizes attack by a single attacking identity. Left: An attacker bypasses the defense by distributing the attack over $N$ attacking identities. Since the defense calculates client-based metrics, each attacker remains below the detection threshold.}
    \label{fig:overview}
\end{figure}

The fallacy of the \gls{SCA} is illustrated in Figure~\ref{fig:overview}. Existing literature often assumes a single attacking entity that generates the entirety of the malicious query load, inevitably causing the client-based metrics to exceed the detection threshold $\tau$ (right). In contrast, the left side demonstrates a distributed attack scenario: Here, the global query volume required to steal the model is partitioned across $N$ attacking identities. Since the defense logic evaluates each client in isolation, the metric for each individual attacker remains below the threshold. As a result, the malicious queries are classified as benign, allowing the attackers to successfully construct a stolen model.

\subsection{Formalization of the Distributed Attacker}

A \gls{MEA} requires a set of queries $Q = \{q_1, \ldots, q_K\}$ to train a substitute model $S$ that resembles the target model $T$. A client-based defense mechanism $D$ monitors a stream of queries $q$ from a client $c$ and triggers an alarm when a metric $M$ exceeds a threshold $\tau$:
\begin{equation}
    D(Q_c) = \begin{cases} 
    1 \; (alarm), & \text{if } M(Q_c) > \tau\\
    0 \; (benign), & \text{otherwise},
    \end{cases}
\end{equation}
where $Q_c \subseteq Q$ is the subset of requests sent by client $c$.

However, a resource-rich attacker (e.g., a state-sponsored actor) may employ a set of clients $C = \{c_1,\ldots, c_N\}$, orchestrated as a botnet or C2 infrastructure. The attacker partitions the necessary set $Q$ into disjoint subsets such that:

\begin{equation}
    Q = \bigcup_{i=1}^{N} Q_{c_i} \quad \text{with} \quad \forall i : |Q_{c_i}| < \epsilon.
\end {equation}

where, $\epsilon$ is a critical threshold below which the statistical significance for the calculation of $M$ is not satisfied (e.g., the minimum sample size required by tests such as Shapiro-Wilk~\cite{shapiro1965analysis}). As long as each individual client $c_i$ operates below this threshold, the overall attack $Q$ remains invisible to the defender, as there is no global correlation of events $D(Q_{c_i})$.

\subsection{Attack Vector 1: Spatial Distribution (round-robin)}

To effectively counteract the \gls{SCA}, we implement a deterministic distribution strategy. Instead of distributing requests randomly, we use round-robin \cite{silberschatz2018operating} scheduling across $K$ clients.
Let $q_t$ be the request at time $t > 0$. The assignment to a client $c_i$ is done by:
\begin{equation}
    Client(q_t) = c_{(t \mod N)}.
\end{equation}
This strategy maximizes the time interval $\Delta t$ between two requests from the same client. For defense-side analyses based on temporal correlations, the data stream of each individual client appears sparse and unsuspicious.
For mechanisms such as \gls{PRADA}, which require a minimum number of requests (e.g., 100 in~\cite{prada}) to achieve statistical significance, this distribution strategy ensures that the detection mechanism is never initiated (see evaluation in Section~\ref{sec:experiments}).

\subsection{Attack Vector 2: Temporal Obfuscation (Traffic Mixing)}

Even if a defense aggregates all incoming requests globally, a statistical approach remains vulnerable to manipulation. Adaptive attackers can mask their signature by mixing attack queries with benign traffic, as illustrated in Algorithm~\ref{alg:traffic_mixing}.

The attacker generates a mixed data stream $Q_{mix} = Q_{attack} \cup Q_{benign}$. The ratio $\lambda = |Q_{attack}| / |Q_{benign}|$ is chosen such that the statistical distribution of the total traffic resembles that of a regular user. In Algorithm~\ref{alg:traffic_mixing}, this is achieved by including a predefined percentage of benign traffic in the stream of attack queries (lines 5-11). The algorithm subsequently alternates between different client identities using round-robin scheduling (lines 4, 7, and 10).
In statistical tests, this implies that the null hypothesis (that the data is benign) can no longer be rejected. A defender attempting to compensate by tightening the thresholds would inevitably cause an increase in \gls{FP}, resulting in a de facto denial of service for legitimate users.

\begin{algorithm}[t]
\caption{Adaptive Traffic Mixing Strategy}
\label{alg:traffic_mixing}
\begin{algorithmic}[1]
    \Require $Q_{attack}$ (Queue attack queries)
    \Require $G_{benign}$ (Generator for benign traffic)
    \Require $\lambda$ (Target mixing ratio, z.B. 0.99)
    \State $Clients \leftarrow \{c_1, \dots, c_N\}$
    \State $Pointer \leftarrow 0$
    \While{$Q_{attack} \neq \emptyset$}
        \State $c_{curr} \leftarrow Clients[Pointer \mod N]$
        \If{$uniform(0,1) < \lambda$}
            \State $q \leftarrow G_{benign}.generate()$
            \State $send(c_{curr}, q)$ \Comment{Send benign traffic}
        \Else
            \State $q \leftarrow Q_{attack}.pop()$
            \State $send(c_{curr}, q)$ \Comment{Send attack query}
        \EndIf
        \State $Pointer \leftarrow Pointer + 1$
        \State $wait(\Delta t)$
    \EndWhile
\end{algorithmic}
\end{algorithm}

The \gls{SCA} represents a conceptual limitation in many proposed \gls{MEA} defenses. The strategies described here show that defenses relying on per-client statistical thresholds fail when attackers distribute their malicious queries over multiple identities and obfuscate statistical patterns by mixing them with benign traffic.
        \section{CerberusAI: a framework for reproducible model stealing research}
\label{sec:framework}

The systematic evaluation of the vulnerabilities outlined in Section~\ref{sec:sca_fallacy} requires an experimental environment that goes beyond a collection of standalone scripts. Existing research often suffers from isolated implementations that are difficult to compare.
To address this gap, we introduce CerberusAI \footnote{\url{https://github.com/lMaxTl/Cerberus-AI}}, a modular open-source framework specifically designed for red teaming of \gls{AI} models and the simulation of complex, distributed attack scenarios.

\subsection{Architecture and Design Principles}

\begin{figure}[tp]
    \centering
    \includegraphics[width=\linewidth]{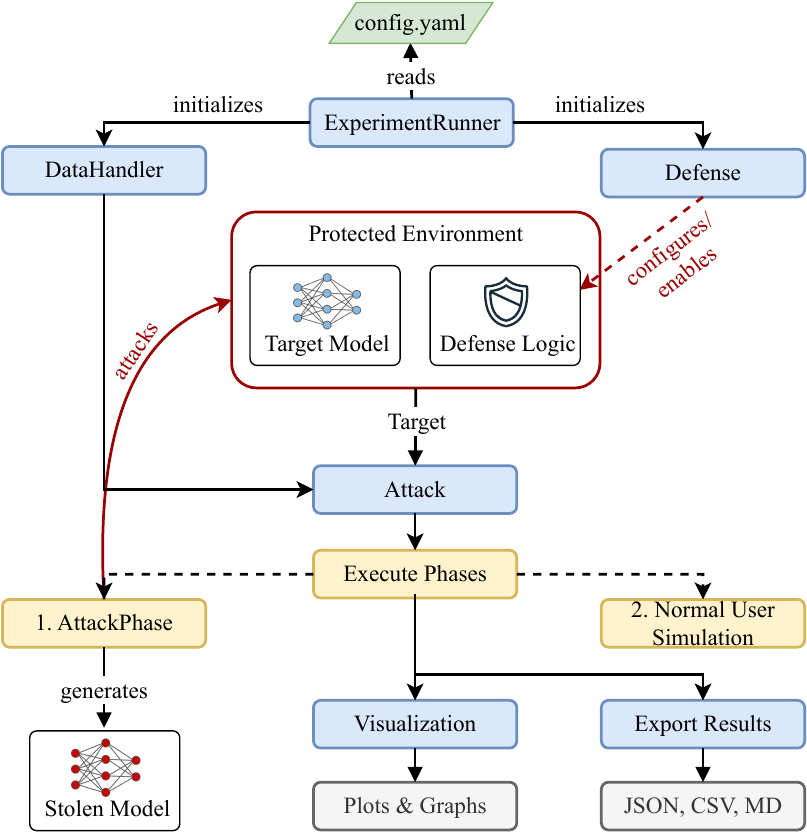} 
    \caption{Abstract flow of an experiment in CerberusAI: From the YAML definition to the orchestration of attack and defense to automated evaluation.}
    \label{fig:framework_flow}
\end{figure}

CerberusAI follows the principle of \emph{separation of concerns} and strictly decouples the attack logic from the defense logic and the target model. The architecture, shown in Fig.~\ref{fig:framework_flow}, is based on three core components:

\begin{itemize}
    \item Declarative configuration: Experiments are defined entirely via \gls{YAML} files. This guarantees reproducibility, as all parameters from the target model and attack strategy to the defense hyperparameters are fixed in a version-controllable file.
    \item Experiment Orchestrator: The \emph{ExperimentRunner} dynamically instantiates the required components using a factory pattern. It encapsulates the target model in a \emph{DefendedModel} wrapper. This wrapper acts as a proxy that first routes each incoming request through the configured defense logic (e.g., \gls{PRADA}) before it reaches the model. This realistically simulates an upstream \gls{API} gateway or \gls{IDS}.
    \item Modular extensibility: A dynamic registry service allows new attacks or defenses to be added without modifying the core code. This enables the rapid integration of new threat vectors.
\end{itemize}

\subsection{Simulation of distributed threats}

A main component of CerberusAI is the \emph{DistributedAttackManager}. This component enables the simulation of botnet-like structures within a controlled laboratory environment.
Instead of instantiating a single attacker, the manager generates $N$ virtual client identities. The attack logic generates the required adversarial queries, which are not sent directly but are instead passed to the manager.

The manager implements various scheduling strategies to technically map the circumvention techniques described earlier in Section~\ref{sec:sca_fallacy}:
\begin{enumerate}
    \item Round-Robin Scheduling: The global budget of attack requests $Q$ is distributed cyclically among the clients $c_1,\ldots,c_N$. This guarantees that no single client exceeds the local detection threshold (e.g., $\epsilon$ requests per minute).
    \item Traffic Mixing Injection: To simulate adaptive attackers, the framework injects configurable benign traffic into the request stream. The parameter \mbox{\texttt{benign\_query\_percentage}} can be used to fine-tune the attack-to-noise ratio $\lambda$ in order to test the sensitivity of global defense mechanisms (see Scenario~4 in Section~\ref{sec:experiments}).
\end{enumerate}

\subsection{Declarative Experiment Definition}

A core goal of CerberusAI is reproducibility. Listing~\ref{lst:config} shows an excerpt from the YAML configuration for a distributed attack. Researchers can specify complex scenarios without the need to implement additional code.

% Ensures that the listing is withing ONE page
\begin{figure}[!tb]
\centering
\noindent\begin{minipage}{\linewidth}
\begin{lstlisting}[language=yaml, caption={Configuration of a distributed attack with traffic mixing.}, label={lst:config}, basicstyle=\ttfamily\small, frame=single]
attack_params:
  attack_type: "prada_attack"
  
  args:
    query_budget: 2500
    epsilon_mifgsm: 0.3
    benign_query_percentage: 0.99
    
  distributed_attack_manager:
    num_clients: 400 
    coordination_strategy: "round_robin"

phases:
  initial_benign_queries:
    type: "benign_simulation"
    num_fixed_queries: 500
    
  attack:
    type: "attack" #executes attack defined with attack_params
    
  post_attack_benign_queries:
    type: "benign_simulation"
\end{lstlisting}
\end{minipage}
 \vspace{-2.0\baselineskip}
 \label{fig:}
\end{figure}

\subsection{Automated Evaluation and Metrics}
CerberusAI integrates a reporting pipeline. Upon completion of an experiment, the framework not only exports logs and tensors but also computes aggregated security metrics:
\begin{itemize}
    \item Attack Success Rate: Performance of the stolen substitute model relative to the target.
    \item Detection Performance: Precision, recall, and F1 score of the defense mechanism.
    \item Cost Analysis: Number of queries required to achieve a successful model extraction.
\end{itemize}
Visualization modules automatically generate graphs showing the detection over time (see Section~\ref{sec:experiments}), enabling in-depth analysis of when and why a defense fails.      
        \section{Experimental Evaluation}
\label{sec:experiments}

In this section, we empirically validate the theoretical assumptions about \gls{SCA}, implemented by CerberusAI, using a set of experiments. The goal is to quantify the effectiveness of the defense mechanism \gls{PRADA} against the defined distributed attack scenarios.

\subsection{Experimental Setup and Parameters}
We use MNIST as the benchmark dataset, as it is well established in \gls{MEA} evaluation and enables direct comparison with the original \gls{PRADA} publication~\cite{prada}.
The target model is a \gls{CNN} achieving 98.4\% test accuracy. We use the \gls{MI-FGSM} algorithm \cite{dong2018boostingadversarialattacksmomentum} as the attack strategy, as it exhibits a substantially higher transfer rate of adversarial examples in preliminary experiments compared to the I-FGSM used in \gls{PRADA}.

The configuration of the experiments is summarized in Table~\ref{tab:exp_config}. A critical detail of the defense is the parameter $N_{min}=50$ (\texttt{min\_set\_size}), which specifies that statistical tests are performed only after an identity has issued at least 50 requests.

\begin{table}[b]
    \centering
    \caption{Configuration parameters of the experiments}
    \label{tab:exp_config}
    \begin{tabular}{l|l}
        \toprule
        \textbf{Parameter} & \textbf{Value / Description} \\
        \midrule
        \textit{Target Model} & CNN (MNIST, 98.4\% Acc) \\
        \textit{Attack Strategy} & MI-FGSM ($\epsilon=0.3$, Iter=60) \\
        \textit{Query Budget} & 2500 Queries (Total) \\
        \textit{Defense} & PRADA ($\delta=0.90$) \\
        \textit{Defense Threshold} & $N_{min}=50$ (Queries per Client) \\
        \textit{Distributed Clients} & 400 (Round-Robin) \\
        \bottomrule
    \end{tabular}
\end{table}

\subsection{Scenarios and Results}
We examine four scenarios to explore the limitations of state-of-the-art defenses. Table~\ref{tab:results} lists the quantitative results (precision, recall, F1 score) observed during experiments.

\begin{table}[b]
    \centering
    \caption{Breakdown of defense performance: While PRADA works in the single-client scenario, the detection rate drops to 0\% in distributed attacks.}
    \label{tab:results}
    \begin{tabular}{lccc}
        \toprule
        \textbf{Scenario} & \textbf{Precision} & \textbf{Recall} & \textbf{F1-score} \\
        \midrule
        1. Baseline (1 client) & 100.0\% & 46.2\% & \textbf{63.2\%} \\
        2. Distributed (400 clients) & - & 0.0\% & \textbf{0.0\%} \\
        3. Global Defense (400 Clients) & 100.0\% & 83.1\% & \textbf{90.8\%} \\
        4. Adaptive (Mixed Traffic) & 1.0\% & 99.3\% & \textbf{2.0\%} \\
        \bottomrule
    \end{tabular}
\end{table}

\begin{figure*}[t]
    \centering
    % --- Line 1: The collapse of the single client assumption ---
    \begin{subfigure}[b]{0.48\textwidth}
        \centering
        \includegraphics[width=\linewidth]{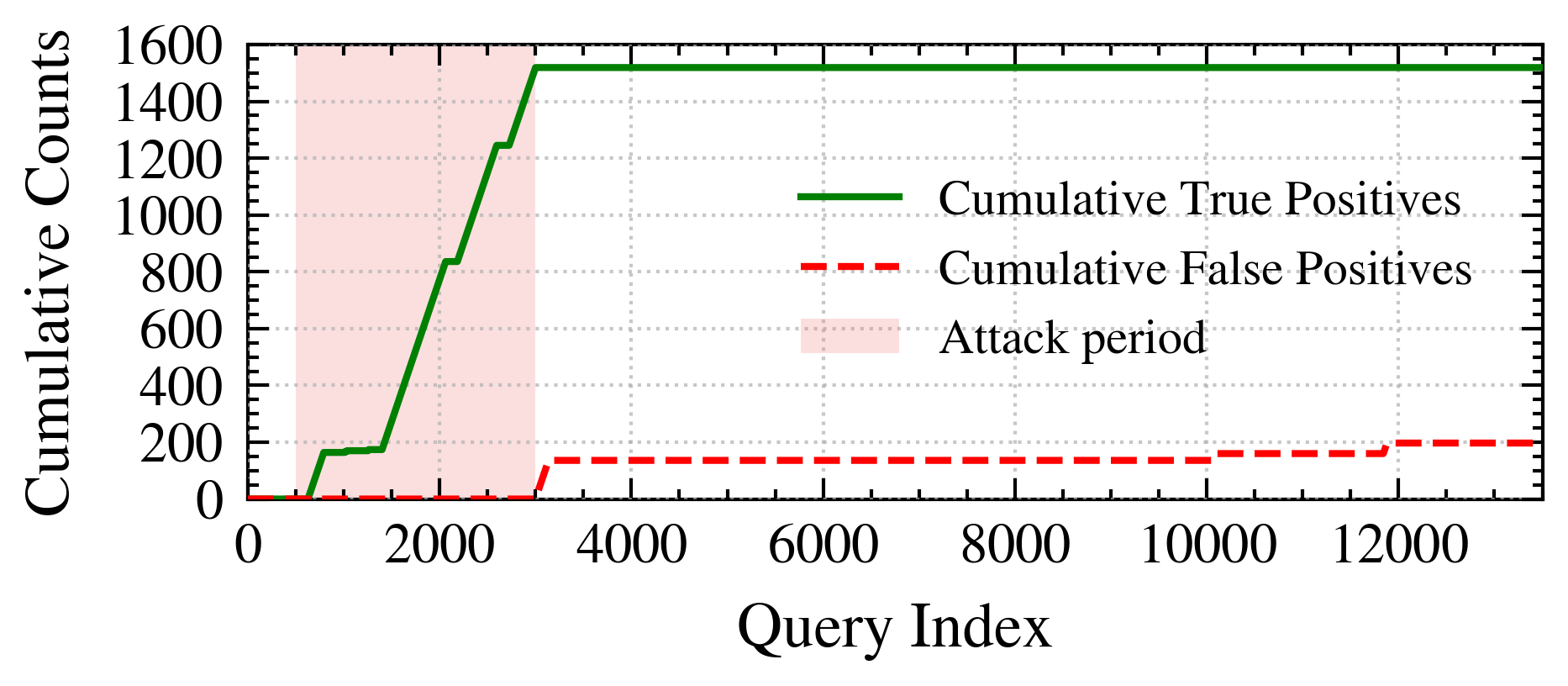} 
        \caption{\textbf{Scenario 1 (Baseline):} Effective detection of a single attacker. The number of true positives (green) increases steadily.}
        \label{fig:res_baseline}
    \end{subfigure}
    \hfill
    \begin{subfigure}[b]{0.48\textwidth}
        \centering
        \includegraphics[width=\linewidth]{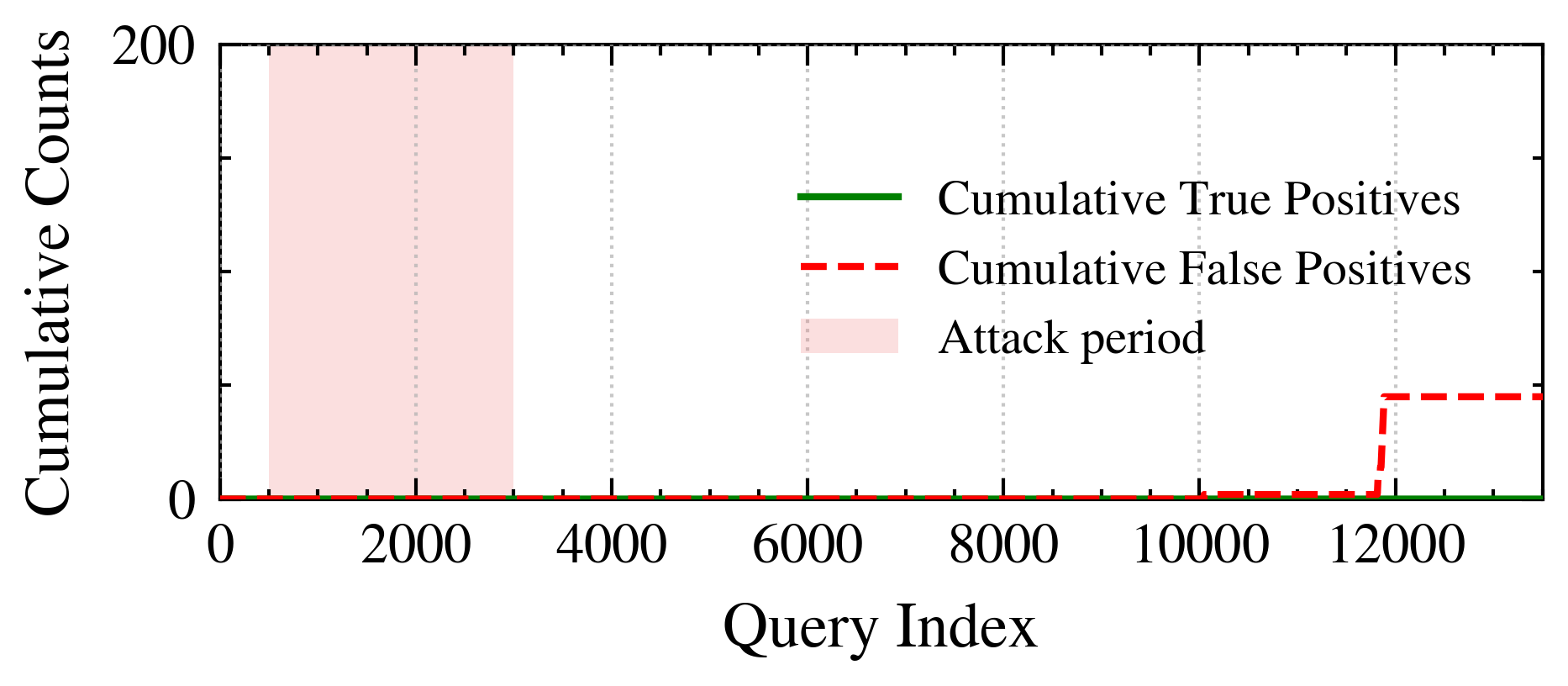}
        \caption{\textbf{Scenario 2 (Distributed):} Total failure of defense. Round-robin keeps every client inconspicuous; only false alarms.}
        \label{fig:res_distributed}
    \end{subfigure}
    
    \vspace{1em} 
    
    \begin{subfigure}[b]{0.48\textwidth}
        \centering
        \includegraphics[width=\linewidth]{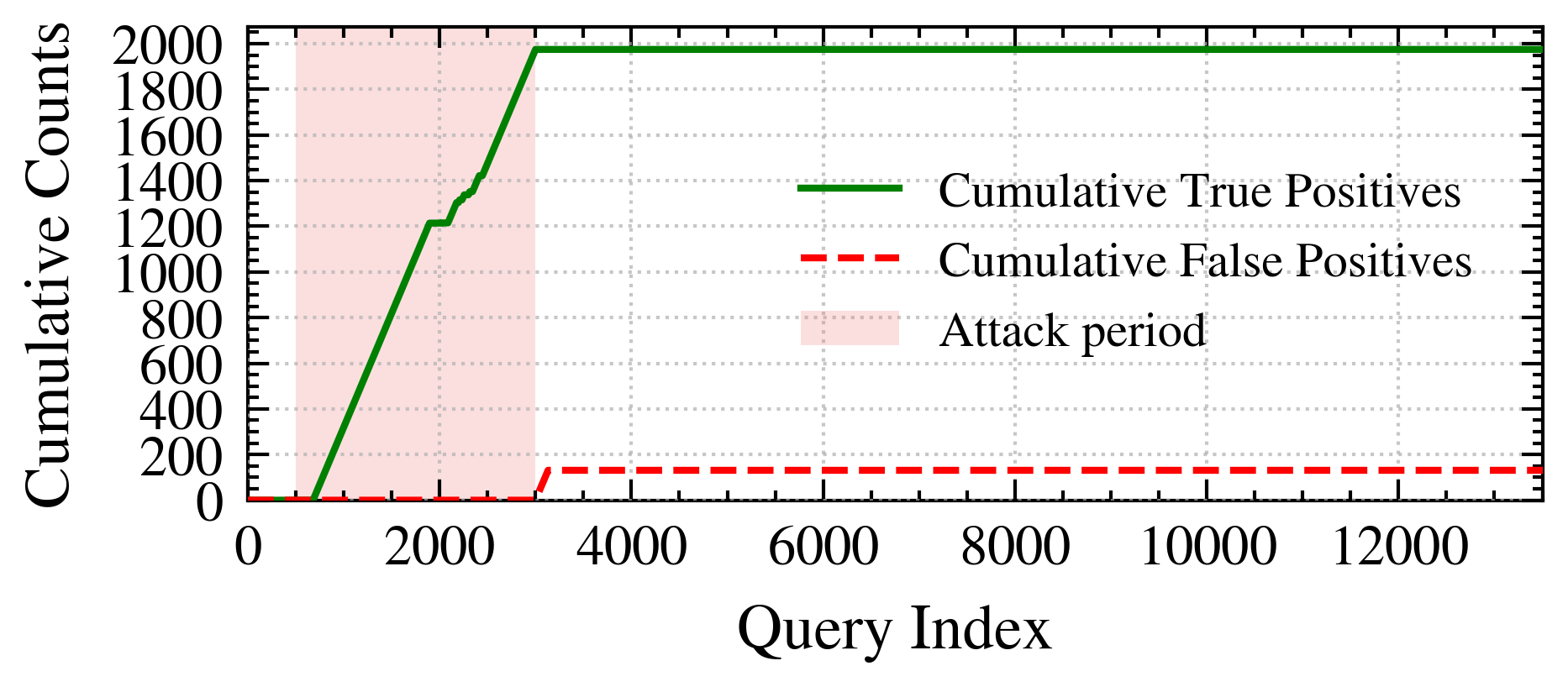}
        \caption{\textbf{Scenario 3 (Global Defense):} Aggregation of all clients makes the attack visible again (similar to baseline).}
        \label{fig:res_global}
    \end{subfigure}
    \hfill
    \begin{subfigure}[b]{0.48\textwidth}
        \centering
        \includegraphics[width=\linewidth]{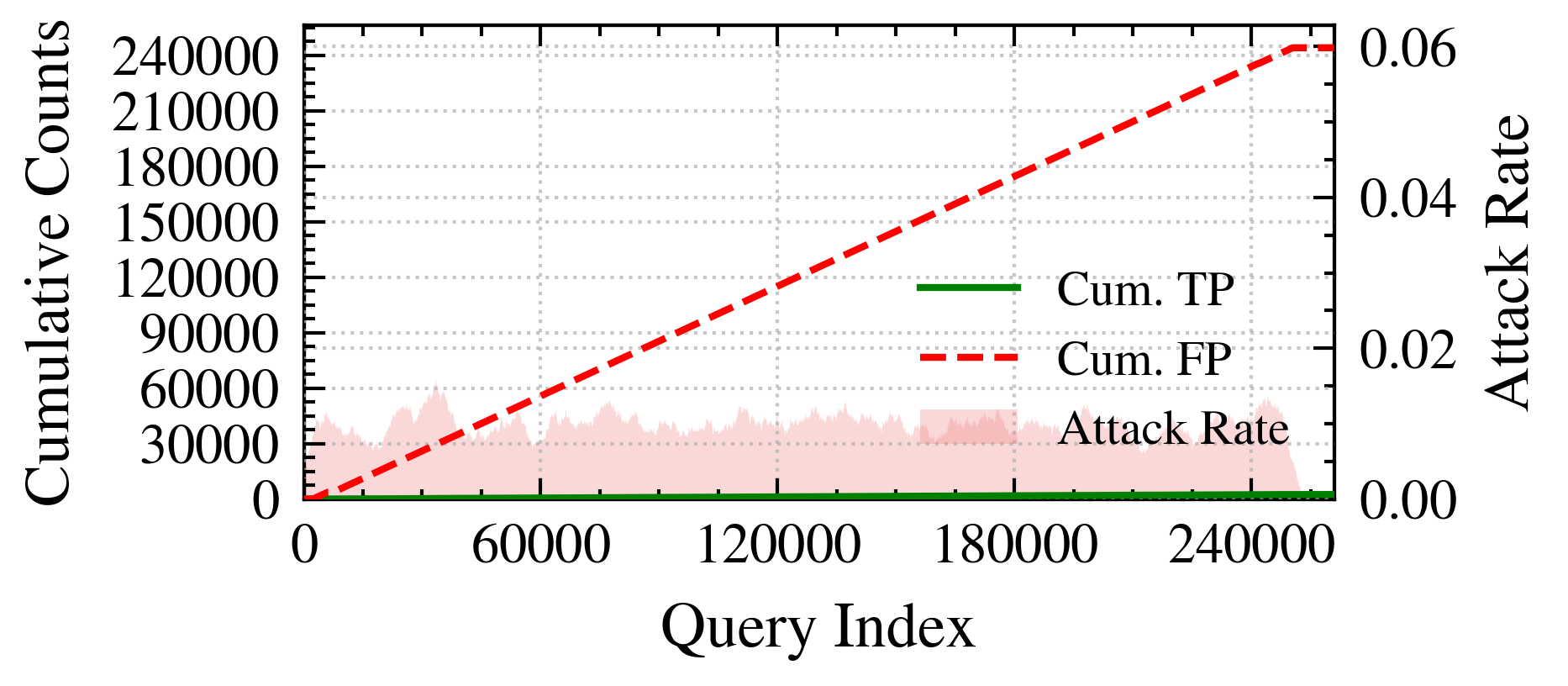}
        \caption{\textbf{Scenario 4 (Adaptive):} Explosion of false positives. Traffic mixing floods the system with false alarms (red).}
        \label{fig:res_adaptive}
    \end{subfigure}
\end{figure*}

\subsubsection{Scenario 1: Baseline (Validation)}
We validate the functionality of the system in the single-client scenario: a single attacker pitted against the original client-based PRADA defense. With an F1 score of 63.2\% and a precision of 100\%, the defense behaves as expected: Attacks are detected as soon as sufficient ($N_{min} = 50$) data points have been observed.

\subsubsection{Scenario 2: The collapse of the SCA}
In this scenario, we distribute the budget of 2,500 requests among 400 clients using round-robin.
The result is consistent with the theoretical analysis in Section~\ref{sec:sca_fallacy}: the F1 score drops to 0.0\%. Since each client sends only about 6-7 requests on average, the threshold $N_{min}=50$ is never reached. Statistical analysis is not initiated for any of the clients.
This provides empirical evidence for the Fallacy of SCA discussed in Section~\ref{sec:sca_fallacy}: A defense that considers identities in isolation is blind to coordinated attacks.

\subsubsection{Scenarios 3 \& 4: Limits of global aggregation}
As a countermeasure, we implement a customized “Global PRADA” variant that aggregates requests across all identities. In scenario~3, this proves effective (F1 score of 90.8\%), as the attack pattern becomes globally visible.

However, scenario~4 (adaptive attacker) demonstrates the fragility of purely statistical approaches: The attackers mix their 2,500 malicious requests with a substantial amount of benign traffic (traffic mixing with a 99\% benign share).
The result is a drop in the F1 score to 2.0\%. Although the recall remains high (the attacks are included in the set), the precision drops to approximately 1\%.
This means that 99 out of 100 alarms are false positives. Such a system is not operationally usable and is equivalent to a denial-of-service of the security infrastructure.

The experiments show that the \gls{SCA} is not a valid basis for security architectures in critical environments. The trivial circumvention by distributed attackers (scenario~2) and the vulnerability of global statistical methods to noise (scenario~4) underscore the need for deterministic, stateful analysis methods that detect attacks based on their semantic content rather than focusing exclusively on their statistical distribution.
        \section{Conclusion}
\label{sec:conclusion}

Securing \gls{AI} models against theft is a critical prerequisite for their deployment in military and security-critical domains. This work provides systematic evidence that the currently prevailing defense doctrine, based on statistical analysis of isolated clients (SCA), is likely to fail against modern, asymmetric threats.
%\subsection{Summary of results}
\glsreset{PRADA}
Our experiments demonstrate that \gls{PRADA}, whose effectiveness is successfully validated under laboratory conditions with a single client, can be trivially bypassed in realistic scenarios. By orchestrating a botnet with 400 clients and a round-robin distribution strategy, the detection performance (F1 score) dropped to 0.0\%. This empirically proves that \gls{SCA} is not a valid assumption for model defense.
In addition, it is shown that naive global aggregation approaches (global defense) help against simple distribution, but can be rendered operationally useless by adaptive attackers who mask their traffic with noise (“traffic mixing”) due to a flood of false positives (precision $< 1\%$).
\glsreset{MEA}
With CerberusAI, we are providing the research community with a tool to close this gap. The framework enables standardized and reproducible research on \glspl{MEA} in distributed settings. Our findings establish two fundamental requirements for the development of resilient defense architectures:

\begin{enumerate}
    \item Defense mechanisms should move away from purely statistical metrics (such as inter-arrival times or distribution matching), as these can be arbitrarily manipulated through distribution and noise.
    \item Promising approaches for future research lie in stateful defenses, which monitor the state of the model globally, and in the semantic analysis of queries. Instead of asking “How fast are the queries coming in?”, future systems should consider the question “What information content do these queries extract in the context of global knowledge?”.
\end{enumerate}

Protecting intellectual property and tactical advantages in algorithmic warfare requires a paradigm shift: away from client-centric anomaly detection toward resilient, identity-independent model monitoring.
        
        \balance
       
        % References
        %\renewcommand{\bibfont}{\smaller}
        \bibliographystyle{IEEEtran}
        \bibliography{references}

        %\tableofcontents
        \balance
       
\end{document}